\documentclass[floatfix,prl,twocolumn,showpacs,amsmath,amssymb,letterpaper]{revtex4}
\usepackage{times}
\usepackage{latexsym}
\usepackage{graphicx}
\usepackage{verbatim,times,bbm}
\usepackage{color}

\newcommand{\Tr}{{\mathrm{Tr}}}

\begin{document}

\title{Capacities of lossy bosonic memory channels}
\author{Cosmo Lupo$^1$, Vittorio Giovannetti$^2$,  and Stefano Mancini$^1$}
\affiliation{ $^1$Dipartimento di Fisica, Universit\`a di Camerino,
via Madonna delle Carceri 9, I-62032 Camerino, Italy \\
$^2$NEST CNR-INFM \& Scuola Normale Superiore, piazza dei Cavalieri
7, I-56126 Pisa, Italy}

\begin{abstract}

We introduce a general model for a lossy bosonic memory channel and
calculate the classical and the quantum capacity, proving that
coherent state encoding is optimal. The use of a proper set
of collective field variables allows to unravel the memory, showing
that the $n$-fold concatenation of the memory channel is unitarily
equivalent to the direct product of $n$ single-mode lossy bosonic
channels.

\end{abstract}


\maketitle

One of the most important problem of quantum information theory is
finding the maxima rates (i.e.\ capacities) at which quantum or
classical information can be transmitted with vanishing error in the
limit of large number of transmitted signals (channel
uses)~\cite{QI}. Earlier works on the subject focused on models
where the noise affecting the communication is assumed to act
independently and identically for each channel use (memoryless
quantum channels). Recently, however, an increasing attention has
been devoted to correlated noise models (memory quantum channels),
see e.g.~\cite{KW2} and Ref.s therein. Memory effects in the
communication may arise when each transmitted signal statistically
depends on both the corresponding and previous inputs. Such scenario
applies when the dynamics of the communication line is characterized
by temporal correlations which extend on timescales which are longer
than the times between consecutive channel uses --- a regime which
can be {\em always} reached  by increasing  the number of
transferred data per second. For instance optical fibers may show
relaxation times or birefringence fluctuations times longer than the
separation between successive light pulses~\cite{exp}. Similar
effects occur in solid state implementations of quantum hardware,
where memory effects  due to low-frequency impurity noise produce
substantial dephasing~\cite{exp1}. Furthermore, moving from the
model introduced in~\cite{GiovMan}, memory noise effects have also
been studied in the contest of many-body quantum systems by relating
their properties  to the correlations of the channel environmental
state~\cite{PLENIO} or by studying the information flow in spin
networks~\cite{SPIN}.

It is generally believed that memory effects should improve the
information transfer of a communication line. However finding
optimal encodings is rather complex and up to date only a limited
examples have been explicitly  solved~\cite{DATTA1,KW2,VJP}. In this
paper we focus on a continuous variable model of quantum memory
channels in which each channel use is described as an independent
bosonic mode. The proposed scheme is characterized by two parameters
which enable us to describes different communication scenarios
ranging from memoryless to intersymbol interference memory
~\cite{BDM}, up to perfect memory configuration~\cite{bowen}. It
effectively mimics the transmission of quantum signals along
attenuating optical fibers characterized by finite relaxation times,
providing the first comprehensive quantum information
characterization of memory effects in these setups. For such model
we exactly calculate the classical and the quantum
capacity~\cite{NOTA0} and prove that coherent state encoding is
optimal. This is accomplished by \emph{unraveling} the memory
effects through a proper choice of encoding and decoding procedures
which transform the quantum channel into a product  of independent
(but not identical) quantum maps. If the channel environment  is in
the vacuum, the capacities can then be computed by using known
results on memoryless  lossy bosonic channels~\cite{Wolf,broadband}
which in the limit of large channel uses provide converging lower
and upper bounds.

\begin{figure}
\centering
\includegraphics[width=0.4\textwidth]{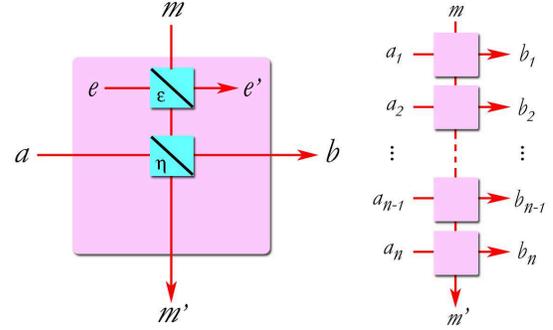}
\caption{Left: a single use of the memory channel (see text for
details). Right: the $n$-fold concatenation of the memory channel:
photons entering in the $k$-th input mode $a_k$ can only emerge in
the output ports $b_{k'}$ with $k'\geqslant k$ (the channel is thus
{\em non-anticipatory}).} \label{memory}
\end{figure}

{\it Channel model:--}  We consider quantum channels described by
assigning a mapping of the form
\begin{eqnarray}\label{equazione1}
\boldsymbol{\Phi}_n(\boldsymbol{\rho}_n) = \Tr_E [
\boldsymbol{U}_{n} (\boldsymbol{\rho}_n \otimes
\boldsymbol{\sigma}_E  )\boldsymbol{U}_{n}^\dag ]\;,
\end{eqnarray}
where $\boldsymbol{\rho}_n$ and
$\boldsymbol{\Phi}_n(\boldsymbol{\rho}_n)$ represent, respectively,
the input and output states of the first $n$ channel uses, and
$\boldsymbol{\sigma}_E$ is the initial state of channel environment
$E$. The latter is composed by a {\em memory kernel} $M$ which
interacts with all inputs, and by a collection $E_1, E_2, \cdots,
E_n$ of {\em local} environments associated with each individual
channel use. Such interactions are described by the unitary
$\boldsymbol{U}_{n}$ which can be taken as a product of identical
terms, i.e.\  $\boldsymbol{U}_{n} =U_n U_{n-1} \cdots U_1$ with
$U_k$ being the interaction between the $k$-th channel input, $E_k$
and $M$. Within this context the channel uses will be described by
an ordered sequence of independent bosonic modes associated with the
input mode operators $\{a_1, a_2, \cdots, a_n\}$. Through the
coupling $\boldsymbol{U}_{n}$ they undergo a damping process that
couples them with the local environments $E_1, E_2, \cdots, E_n$ and
the memory kernel $M$ (also  described by a collection of mode
operators $\{e_1, e_2, \cdots, e_n\}$ and $m_1$). Memory effects
arise when the photons lost by the $k$-th channels mix with the
environmental mode $e_{k+1}$ of the subsequent channel use.
Specifically the evolution of $k$-th input is obtained by a
concatenation of two beam-splitter transformations, the first with
transmissivity $\epsilon$ and the second with transmissivity $\eta$
(see Fig.~\ref{memory}, left). In the Heisenberg-picture this is
defined by the  identities
\begin{eqnarray}
m_k' & = & \sqrt{\epsilon\eta} \, m_k + \sqrt{1-\eta} \, a_k + \sqrt{\eta(1-\epsilon)} \, e_k, \nonumber\\
b_k & = & - \sqrt{\epsilon(1-\eta)} \,m_k +  \sqrt{\eta} \, a_k- \sqrt{(1-\epsilon)(1-\eta)} \, e_k, \nonumber \\
e_k' & = & -\sqrt{1-\epsilon} \, m_k + \sqrt{\epsilon} \, e_k, \label{equaz1}
\end{eqnarray}
where $m_k' := U_k^\dag m_k U_k$, $b_k := U_k^\dag a_k U_k$, and
$e_k' := U_k^\dag e_k U_k$ describe the  outgoing modes of the model
(in particular the $b_k$'s are associated with the receiver
signals). The resulting input/output mapping is finally obtained by
a $n$-fold concatenation of Eq.s~(\ref{equaz1}) where, for each $k$,
we identify the mode $m_{k+1}$ with $m_{k}'$ (see Fig.~\ref{memory},
right). This yields a {\em non-anticipatory}~\cite{GALLA} channel
where  a given  input can only influence subsequent channel outputs
(i.e.\ for each $k$, $b_k$ depends only upon the $a_{k'}$'s with
$k'\leqslant k$). The transmissivity $\epsilon$ plays the role of a
memory parameter. In particular the model reduces to a memoryless
channel~\cite{broadband} for $\epsilon=0$ (the input $a_k$ only
influences the output $b_k$), and to a channel with perfect memory
\cite{bowen} for $\epsilon=1$ (all $a_k$ interacts {\em only} with
the memory mode $m_1$). Intermediate configurations are associated
with values $\epsilon \in ]0,1[$ and correspond to intersymbol
interference channels where the previous input states affect the
action of the channel on the current input~\cite{BDM}. Of particular
interest is also the case $\eta=0$ where $\boldsymbol{\Phi}_n$
describes  a {\em quantum shift} channel~\cite{BDM}, where each
input state is replaced by the previous one.

When dealing with memory channels, four different cases can be
distinguished depending whom the memory mode is assigned
to~\cite{KW2}. Specifically the initial and final state of the
memory can be under the control of the sender of the message ($A$),
the receiver ($B$) or the environment ($E$). The four possible $XY$
setups are denoted: $XY=AB$ (initial memory to $A$ and final memory
to $B$), $XY=AE$, $XY=EB$, $XY=EE$. These different scenarios
typically lead to different values of the channel capacity but, at
least for finite dimensional system, they coincide if the channel is
forgetful~\cite{KW2}. To make the notation homogeneous we thus
define: $a_0 := m_1$ and $b_{n+1} := m'_n$ if $XY=AB$; $a_0 := m_1$
and $e'_{n+1} := m'_n$ if $XY=AE$; $e_0 := m_1$ and $b_{n+1} :=
m'_n$ if $XY=EB$; $e_0 := m_1$ and $e'_{n+1} := m'_n$ if $XY=EE$.

\begin{figure}[t]
\centering
\includegraphics[width=0.45\textwidth]{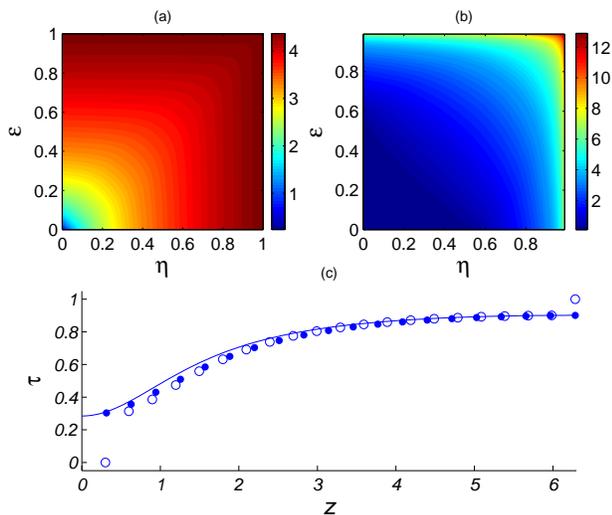}
\caption{(Color online.) (a) Contour plot of the classical capacity
 $C$ as function of $\eta$ and
$\epsilon$ for $N=8$. (b) Contour plot of the unconstrained quantum capacity
$Q_\infty$, in the $XY=EE,AE$ setups, as function of $\eta$ and
$\epsilon$ (it diverges logarithmically
at $\eta=1$ and $\epsilon=1$). (c) Distribution of the effective
transmissivities $\tau^{XY}_k$ for $\eta=0.7$, $\epsilon=0.3$: the
solid line shows the asymptotic distribution computed from
Eq.~(\ref{spectrum}), the dots and the circles respectively show the
distribution of transmissivities $\tau_k^{EE}$ and $\tau_k^{AB}$ for $n=20$.}
\label{capacities}
\end{figure}

With the above choices the output modes of the receiver can then be
expressed in the following compact form
\begin{equation}\label{equazione2}
b_k = \boldsymbol{U}_{n}^\dag \; a_k \; \boldsymbol{U}_{n}= A_k^{XY}
+ E_k^{XY}\;,
\end{equation}
with $A_k^{XY}$ and $E_k^{XY}$ being, respectively, field operators
formed by linear combination of  the field modes $a_{k'}$ and
$e_{k'}$ with $k'\leqslant k$ [The explicit expressions can be easily derived
from~Eq.~(\ref{equaz1}) but are not reported here because they are
rather cumbersome]. The $A_k^{XY}$ commute with the $E_k^{XY}$
together with their hermitian conjugates. Furthermore they satisfy
the following commutation relations:
\begin{eqnarray}
{[} A_k^{XY} , {A_{k'}^{XY}}^\dag {]} =  M_{kk'}^{XY}, \quad
{[} E_k^{XY} , {E_{k'}^{XY}}^\dag {]} = \delta_{kk'} -  M_{kk'}^{XY}, \nonumber
\end{eqnarray}
with $\delta_{kk'}$ being the Kronecker delta and $M^{XY}$ being a
symmetric, positive real matrix which satisfies the condition
$\openone \geqslant M^{XY}$. For example the $n \times n$ matrix
$M^{EE}$ has elements
\begin{eqnarray} \nonumber
M_{kk'}^{EE} & = & \delta_{kk'} - (1-\eta_{\min\{k,k'\}}) \sqrt{\epsilon\eta}^{|k-k'|},
\label{equaz4}
\end{eqnarray}
with $\eta_k := \eta +({1-(\epsilon\eta)^{k-1}}) \tfrac{\epsilon
(1-\eta)^2}{1-\epsilon\eta}$. Analogous expressions hold for
$XY=AB$, $AE$ and $EB$  which only differ by terms which in the
limit of $n\rightarrow\infty$ can be neglected. Indeed, by varying
$n$, the $M^{XY}$ form a sequence of matrices of increasing
dimensions  which (independently from $XY$) are {\em asymptotically
equivalent}~\cite{toeplitz} to the Toeplitz matrix $M^{(\infty)}$ of
elements
\begin{equation}\label{matrix-infty}
M_{kk'}^{(\infty)} := \delta_{kk'} - (1-\eta^{(\infty)})
\sqrt{\epsilon\eta}^{|k-k'|},
\end{equation}
with $\eta^{(\infty)} := \lim_{k\rightarrow \infty } \eta_k = \eta +
\frac{\epsilon (1-\eta)^2}{1-\epsilon\eta}$. Similarly the
asymptotic distribution of the eigenvalues $\tau^{XY}_k$ of $M^{XY}$
can be computed by performing the Fourier transform of the matrix
$M^{(\infty)}$~\cite{toeplitz}. Defining $z:=2\pi k/n$ and taking
$n\rightarrow \infty$ this  gives the nondecreasing function
\begin{equation}\label{spectrum} \tau(z) =
\frac{\epsilon+\eta-2\sqrt{\epsilon\eta}\cos{(z/2)}}{1+\epsilon\eta-2\sqrt{\epsilon\eta}\cos{(z/2)}}
= \left|
\frac{\sqrt{\epsilon}-\sqrt{\eta}\,e^{iz/2}}{1-\sqrt{\epsilon\eta}\,e^{iz/2}}
\right|^2\;,
\end{equation}
which is plotted in Fig.~\ref{capacities}(c). According to the
Szeg\"{o} theorem~\cite{toeplitz} the asymptotic average of any
smooth function $F$ of the eigenvalues of $M^{XY}$ can then be
computed by the formula
\begin{equation}\label{szego}
\lim_{n\rightarrow\infty} \frac{1}{n} \sum_{k}F(\tau^{XY}_k)
= \int_0^{2\pi} \frac{dz}{2\pi} F(\tau(z)),
\end{equation}
which is explicitly non dependent upon $XY$.

{\it Unraveling the memory:--} We show that the memory effects can
be unraveled by introducing a proper set of collective coordinates.
To do so we introduce the (real) orthogonal matrix $O^{XY}$ which
diagonalizes the matrix $M^{XY}$ (it exists since the latter is real
symmetric), i.e.\ $\sum_{r,r'} O^{XY}_{kr} M^{XY}_{rr'}
O^{XY}_{k'r'} = \delta_{kk'} \tau^{XY}_k$  (here the $\tau^{XY}_k \in
[0,1]$ are intended to be arranged in nondecreasing
order).

 Let us define the following sets of operators $\mathrm{b}_k
:= \sum_{k'} O_{kk'}^{XY} b_{k'}$, $\mathrm{a}_k := \sum_{k'}
O_{kk'}^{XY} A_{k'}^{XY}/\sqrt{\tau^{XY}_k}$, $\mathrm{e}_k :=
\sum_{k'} O_{kk'}^{XY} E_{k'}^{XY}/\sqrt{1-\tau^{XY}_k}$. By
construction they satisfy canonical commutation relations, moreover
it is easy to show that they obey the following transformations
\begin{equation} \label{eqn555}
\mathrm{b}_k  = \boldsymbol{U}_{n}^\dag  \mathrm{a}_k
\boldsymbol{U}_{n}= \sqrt{\tau^{XY}_k} \;  \mathrm{a}_k +
\sqrt{1-\tau^{XY}_k} \; \mathrm{e}_k \;.
\end{equation}
We denote by $W_A$, $V_B$, $T_E$
 the canonical
unitaries~\cite{HOLEVOBOOK} that implement the transformations $a_k
\rightarrow \mathrm{a}_k = W_A^\dag a_k W_A$, $b_k \rightarrow
\mathrm{b}_k = V_B^\dag b_k V_B$ and $e_k \rightarrow \mathrm{e}_k =
T_E^\dag e_k T_E$. We have shown that the channel
$\boldsymbol{\Phi}_n$ is unitarily equivalent to the map
\begin{eqnarray} \label{nuova}
\boldsymbol{\Phi}^\prime_n(\boldsymbol{\rho}_n)= \Tr_E [
\boldsymbol{U}_{n}^{\prime} ( \boldsymbol{\rho}_n \otimes
\boldsymbol{\sigma}_E^\prime )(\boldsymbol{U}_{n}^{\prime})^\dag
]\;,
\end{eqnarray}
with $\boldsymbol{\sigma}_E^\prime := T_E^\dag \boldsymbol{\sigma}_E
T_E$, and where the unitary transformation
$\boldsymbol{U}_{n}^{\prime} := V_A \boldsymbol{U}_{n} (W_A \otimes
T_E)$ induces the beam-splitter transformations in
(\ref{eqn555})~\cite{NOTA33}. Formally, the unitary equivalence
reads $\boldsymbol{\Phi}^\prime_n(\boldsymbol{\rho}_n) =  V_A \;
{\boldsymbol{\Phi}}_n( W_A^\dag  \boldsymbol{\rho}_n W_A) V_A^\dag$,
i.e.\ we can treat the output states of ${\boldsymbol{\Phi}}_n$ as
output of ${\boldsymbol{\Phi}'}_n$ by first counter-rotating the
input $\boldsymbol{\rho}_n$ by $W_A$ (coding transformation) and
then by rotating the output by $V_A$ (decoding)\cite{GiovMan}.
Assuming then $\boldsymbol{\sigma}_E$ to be the vacuum state, we
have $\boldsymbol{\sigma}_E^\prime = \boldsymbol{\sigma}_E$ and the
map~(\ref{nuova}) can be written as a direct product of a collection
of independent  lossy bosonic channels, i.e.\
\begin{equation}\label{product}
\boldsymbol{\Phi}^\prime_n = \bigotimes_{k} {\Phi}_k,
\end{equation}
with ${\Phi}_k$ being a single-mode lossy bosonic channel with
effective transmissivity $\tau^{XY}_k$.

{\it Classical capacity:--} Equation~(\ref{product}) suggests that
we can compute the classical capacity of $\boldsymbol{\Phi}_n$ by
applying the results of Ref.~\cite{broadband} on memoryless
multi-mode lossy channel. To do so however, we have first to deal
with the fact that the single-mode channels forming
$\boldsymbol{\Phi}^\prime_n$ are not necessarily identical (indeed,
for finite $n$ their transmissivities $\tau_k^{XY}$ can be rather
different from each other). Therefore the map~(\ref{product}) is not
memoryless in the strict sense. To cope with this problem we will
construct two collections of memoryless  multi-mode channels which
upper and lower bound the capacity of $\boldsymbol{\Phi}^\prime_n$
(and thus of ${\boldsymbol{\Phi}}_n$), and use the asymptotic
properties of the distribution~(\ref{spectrum}) to show that for
large $n$ they converge toward the same quantity.

First, as usually done when dealing with bosonic
channels~\cite{HolevoWerner}, we introduce a constraint on the
average photon number per mode of the inputs signals. This yields
the inequality $\frac{1}{n} \sum_k \Tr [ a_k^\dag a_k
\boldsymbol{\rho}_n ] \le N$, which is preserved by the encoding
transformation $\boldsymbol{\rho}_n \rightarrow  W^\dag_A
\boldsymbol{\rho}_n W_A$  of Eq.~(\ref{nuova}) due to the fact that
$W_A$ is  a canonical unitary, i.e.\ $ \frac{1}{n} \sum_k  \Tr [
a_k^\dag a_k W^\dag_A \boldsymbol{\rho}_n W_A ] \leqslant N$. For any
$n$, we then group the single-mode channels of Eq.~(\ref{product})
in $J$ blocks, each of size $\ell=n/J$. At the boundary of the
$j$-th block the minimum and maximum limits of the effective
transmissivities are defined as
\begin{eqnarray}
\underline{\tau}^{XY}_j :=  \liminf_{n\to\infty} \tau^{XY}_{(j-1)n/J+1}, \;\;
\overline{\tau}^{XY}_j :=  \limsup_{n\to\infty}
\tau^{XY}_{jn/J}.
\end{eqnarray}
Hence, recalling that the $\tau_k^{XY}$'s are in nondecreasing order,
we may notice that  for any $\delta > 0$ and for sufficiently large
$\ell$
\begin{equation}
\underline{\tau}^{XY}_j - \delta < \tau^{XY}_{(j-1)\ell+k} <
\overline{\tau}^{XY}_j + \delta, \quad k=1,\dots \ell. \label{efeq1}
\end{equation}
For each $J$, we are thus led to define two new sets of  memoryless
multi-mode lossy channels characterized, respectively, by the two
sets of transmissivities $\{\underline{\tau}^{XY}_j\}_{j=1,\dots
J}$, and $\{\overline{\tau}^{XY}_j\}_{j=1,\dots J}$. Taking the
limit $\ell\to\infty$ while keeping $J$ constant, their capacities
can be computed as in Ref.~\cite{broadband} yielding
\begin{equation}\label{fff}
\underline{C}=\frac{1}{J} \sum_{j=1}^J g(\underline{\tau}^{XY}_j
\underline{N}_j) \;, \qquad \overline{C}= \frac{1}{J} \sum_{j=1}^J
g(\overline{\tau}^{XY}_j \overline{N}_j)\;,
\end{equation}
where $g(x) := (x+1)\log_2{(x+1)} - x\log_2{x}$~\cite{NOTE3}. The
optimal  photon numbers $\underline{N}_j$ and $\overline{N}_j$ are
chosen in order to satisfy the energy constraint (\ref{efeq1}) and
to guarantee the maximum values of  $\underline{C}$ and
$\overline{C}$ respectively. Furthermore Eq.~(\ref{efeq1}) shows
that, one by one, each lossy channel entering the rhs of
Eq.~(\ref{product}) can be lower or upper bounded by the
corresponding channel of the two sets (this is a trivial consequence
of the fact that a lossy channel can simulate those of smaller
transmissivity). Therefore the capacity of $\boldsymbol{\Phi}_n$ can
be  bounded by the capacities $\underline{C}$ and $\overline{C}$ of
Eq.~(\ref{fff}), i.e.\
\begin{equation}
\frac{1}{J} \sum_{j=1}^J g(\underline{\tau}^{XY}_j \underline{N}_j)
\le C \le \frac{1}{J} \sum_{j=1}^J g(\overline{\tau}^{XY}_j
\overline{N}_j),
\end{equation}
which applies for all $J$ and for all $XY$. Taking the limit
$J\to\infty$ and applying (\ref{szego}) we notice that the two
bounds converge to the same quantity. Therefore we conclude that
\begin{equation}\label{classical}
C = \int_0^{2\pi} \frac{dz}{2\pi} g(\tau(z) N(z)),
\end{equation}
with $N(z)$ being the optimal photon number distribution.
Following~\cite{broadband} it can be computed as $N(z) =
[\tau(z)(2^{L/\tau(z)}-1)]^{-1}$ where $L$ is a Lagrange multiplier
whose value is determined by the implicit integral equation
$\int_0^{2\pi} \frac{dz}{2\pi} N(z) = N$, which enforces the input
energy constraint. In some limiting cases Eq.~(\ref{classical})
admits a close analytical solution. For instance in the memoryless
configuration $\epsilon=0$, we get $\tau(z) = \eta$, $N(z)=N$ and
thus correctly $C=g(\eta N)$~\cite{broadband}. Vice-versa for
$\eta=1$ (noiseless channel) or $\epsilon=1$ (perfect memory
channel) we have $\tau(z)=1$, $N(z)=N$ and thus $C=g(N)$ (perfect
transfer). Finally for $\eta=0$ (quantum shift channel) we get
$\tau(z)=\epsilon$, $N(z)=N$ and thus $C=g(\epsilon N)$. For generic
values of the parameters the resulting expression can be numerically
evaluated, showing an increase of $C$ for increasing memory
$\epsilon$ --- see Fig.~\ref{capacities}(a).

{\it Quantum capacity:--} We proceed as  in the previous case and
use the results of Ref.~\cite{Wolf} for the quantum capacity on
memoryless lossy channels  to produce the following bounds on the
quantum capacity of $\boldsymbol{\Phi}_n$
\begin{eqnarray}
\frac{1}{J} \sum_{j=1}^J
{q}(\underline{\tau}^{XY}_j,\underline{N}_j) \le Q \le
\frac{1}{J} \sum_{j=1}^J
{q}(\overline{\tau}^{XY}_j,\overline{N}_j),
\end{eqnarray}
which holds for all $J$. Here ${q}(\tau,N) = \max \{ 0, g(\tau N) -
g((1-\tau)N) \}$ is the maximal coherent
information~\cite{HolevoWerner} and the optimal photon number
distributions $\underline{N}_j$, $\overline{N}_j$ can be  computed
as in Ref.~\cite{broadband}. Finally we take the limit
$J\to\infty$ applying Eq.~(\ref{szego}) to the
function~${q}(\tau,N)$, yielding $Q = \int_0^{2\pi}
\frac{dz}{2\pi} {q}(\tau(z), N(z))$,
with the optimal photon number distribution $N(z)$ to be computed
numerically.

A little thought leads to recognize that in the $XY=EB,AB$ setups,
where the output memory is assigned to Bob, there is at least one
mode which is transmitted with unit efficiency for any value of
$\epsilon$. In the case of unconstrained input energy this leads to
infinite quantum capacity, implying that the limits $n\to\infty$ and
$N\to\infty$ do not commute. A numerical evaluation indicates that,
in the $XY=EE,AE$ setups, the distribution of the transmissivities
converges uniformly to the function in (\ref{spectrum}). This
implies that the formula (\ref{szego}) can be applied even in the
unconstrained case, yielding $Q_\infty = \int_0^{2\pi}
\frac{dz}{2\pi} q(\tau(z))$, where $q(x) := \max \{ 0, \log_2{x} -
\log_2{(1-x)} \}$ -- see Fig.~\ref{capacities}(b).

{\it Conclusions:--} We have computed the capacities of a broad
class of lossy bosonic memory channels without invoking their
forgetfulness. Proving that the channel~(\ref{equazione1}) is
forgetful requires to show that in the limit $n\to\infty$ the final
state of the memory $M$ (i.e.\ the state associated with the mode
$m'_n$) is independent, in the sense specified in Ref.~\cite{KW2},
on the memory initialization. A simple heuristic argument suggests
that this is the case. The argument goes as follows: a photon
entering from the input port $m_1$ of the setup has only an
exponentially decay probability $({\epsilon\eta})^n$ of emerging
from  the $m_n'$ output port (this is the probability of passing
through the sequence of $n$ beam-splitters of of Fig.~\ref{memory}).
Consequently the contribution of $m_1$ to the output state $m_n'$ is
negligible for large values of $n$. If one restricts the analysis to
Gaussian inputs with bounded energy this observation can be
formalized in a rigorous proof. However generalizing it to non
Gaussian inputs is problematic due to the infinite dimension of the
associated Hilbert spaces~\cite{nota}. Moreover, our results on the
quantum capacity suggest that the channel is not forgetful if the
input energy is unconstrained.

We conclude by noticing that the optimal encoding strategy for the
memoryless channels which bound $\boldsymbol{\Phi}_n$ make use of
coherent states~\cite{broadband}. Since the latter are preserved by
the encoding transformation $W_A$ our results prove, as a byproduct,
the optimality of coherent state encoding for the memory channel.

V.G.\ acknowledges Centro De Giorgi of SNS for financial support.
C.L.\ and S.M.\ acknowledge the financial support of the EU under
the FET-open grant agreement CORNER, number FP7-ICT-213681.


\begin{thebibliography}{99}
\bibitem{NOTA0} Similar results can be also derived for their entangled assisted counterparts~\cite{QI}.
\bibitem{NOTE3} The new sets can be treated as memoryless channels, as for
$n\rightarrow \infty$  we have $\ell =n/J\rightarrow \infty$ copies
of their transmissivities.
\bibitem{nota} Also we notice that the notion of forgetfulness
as introduced in \cite{KW2} cannot be directly applied to the
infinite dimensional case, and a proper extension would be needed.
\end{thebibliography}
\end{document}